\documentclass[a4paper, 12pt]{article}

\catcode`\ç\active\def ç{{\c c}} \catcode`\Ç\active\def Ç{{\c C}}
\catcode`\ð\active\def ð{{\u g}} \catcode`\Ð\active\def Ð{{\u G}}
\catcode`\ý\active\def ý{{\i}} \catcode`\Ý\active\def Ý{{\. I}}
\catcode`\ö\active\def ö{{\" o}} \catcode`\Ö\active\def Ö{{\" O}}
\catcode`\þ\active\def þ{{\c s}} \catcode`\Þ\active\def Þ{{\c S}}
\catcode`\ü\active\def ü{{\" u}} \catcode`\Ü\active\def Ü{{\" U}}

\begin{document}
\setlength{\baselineskip}{1.3\baselineskip}
\begin{center}
\textbf{A HASH OF HASH FUNCTIONS} \\[0.25in]
TÜRKER ÖZSARI \\[0.15in]
Departments of Mathematics and Computer Engineering, \\[0pt]
\smallskip Koç University, Sarýyer, Ýstanbul, 34450 \\[0pt]
\smallskip tozsari@ku.edu.tr \\[0.15in]
\today
\\[0.5in]
\textbf{Abstract}
\end{center}

\noindent {In this paper, we present a general review of hash
functions in a cryptographic sense. We give special emphasis on
some particular topics such as cipher block chaining message
authentication code (CBC MAC) and its variants.  This paper also broadens the information given in [1], by
including more details on block-cipher based hash functions and
security of different hash schemes.}
\newpage
\tableofcontents
\newpage
\section{Introduction}
In dictionary, \textit{hash} is defined as 'to chop into small
pieces' or 'to muddle, mess up', [2]. The mathematics,
particularly cryptography implicitly combines these two meanings
of hashing to define the term, hash function.  The reason is
that, in cryptography a hash function is the name of a process of
taking an input, a message, and messing it up by an algorithm,
and finally producing a smaller output, message digest (hash
value), compared to the input.  Hash functions are constructed
with the design intent that a hash value should be like a
fingerprint of the message.  For a hash value, to be like a
fingerprint means that two randomly chosen messages would have
the same hash value with sufficiently small probability.  In
other words, hash values should be compressed representatives of the messages they correspond.

These properties of hash functions provide that they can be used for data integrity and message authentication.
A hash function can be used for data integrity as follows.  Suppose that there is a sender (\textit{S}) who will send a message \textit{M} to a receiver (\textit{R}).  \textit{S} first computes the hash value \textit{h(M)} of the message in question with a hash function \textit{h}.  Then sends \textit{M} together with \textit{h(M)} to \textit{R}.  When \textit{R} gets these, he recomputes the hash value of the possibly modifed message \textit{M} and compares it with the original
hash value \textit{h(M)}.  If these two are equal, then \textit{R} believes that \textit{M} is not changed.\footnote{Here,
we assume that \textit{h(M)} is not affected during transmission.}  If \textit{S} and \textit{R} also want the authentication of the message sent in this transmission, \textit{S} can compute his signature by using an encryption algorithm \textit{$E_K(.)$} with key \textit{K}.  Here, \textit{S} encrypts the hash value \textit{h(M)} as \textit{$C=E_K(H(M))$} and sends \textit{M} together with \textit{C} instead of \textit{h(M)} to \textit{R}.  When \textit{R} gets these, by using a verification algorithm \textit{$V_K(.)$}, he does or does not believe that \textit{M} was sent by \textit{R} depending on the result of verification.
\newpage
\section{A Formal Perspective}
It is now time to explore hash functions more formally.
\subsection{Fundamental Definitions and Results}
A very general mathematical definition of a hash function is the following.

\noindent \textbf{Definition 2.1.1} A \textbf{hash function} is a function \textit{$h:\{0,1\}^*\rightarrow\{0,1\}^n$} for a fixed positive integer $n$ and with the property that $h(x)$ is easy to compute for all $x\in\{0,1\}^*$ for any person.

Although the above definition refers to the unkeyed hash functions, there are also keyed hash functions where
a hash value is determined by two inputs, a message and a secret key.  In [1], a keyed hash function is defined as follows.

\noindent \textbf{Definition 2.1.2} A \textbf{keyed hash function} is a function \textit{$h:\{0,1\}^{\kappa}\times\{0,1\}^*\rightarrow\{0,1\}^n$} for fixed positive integers $n$ and $\kappa$ if it satisfies the following properties:

1. $h$ is public and $h(k,x)$ is easy to compute for all $x\in\{0,1\}^*$ and $k\in\{0,1\}^{\kappa}$.

2. Without knowing $k$, it is hard to find x when $h(k,x)$ is given, it is also hard to find two messages $x$ and $x'$ with
$h(k,x)=h(k,x')$.

3. Given zero or more $(x,h(k,x))$ pairs it is hard to find $k$.  This property is called as key non-recovery.

4. Without knowing $k$, it is hard to compute $h(k,x)$ for any $x$ even there is a large set of known $(x_i,h(k,x_i))$ pairs, of course for $x \neq x_i$.

There are several approaches of classifying hash functions.  For example, one grouping method is to divide these functions such as keyed and unkeyed hash functions as told above.  A second approach is to divide them as block-cipher-based and non-block-cipher-based hash functions.  Another classification is to examine these functions based on the specific requirements they have.  Whatever the classification is, hash functions depending on the specific application they are used for, should have some significant properties for security reasons.  Some of these properties are as follows.

1. \textbf{Preimage Resistance}: Given $y\in\{0,1\}^n$, it is hard to find any $x\in\{0,1\}^* \ni h(x)=y$.

2. \textbf{2nd-Preimage Resistance}: Given $x$, it is hard to find any $x'\neq x\ni h(x)=h(x')$.

3. \textbf{Collision Resistance}: It is hard to find any $x$ and $x'\neq x\ni h(x)=h(x')$.

The most important class of unkeyed hash functions are the manipulation detection codes (MDCs) which are used for data integrity.  We can divide MDCs into two groups one of which is the one way hash functions (OWHFs) and the other one is the collision resistant hash functions (CRHFs), both are defined as follows, [3].

\noindent \textbf{Definition 2.1.3} A \textbf{one way hash function} is a hash function with properties preimage resistance and second preimage resistance.

\noindent \textbf{Definition 2.1.4} A \textbf{collision resistant hash function} is a hash function with properties 2nd-preimage resistance and collision resistance.

Some sources may define one way hash function without the property 2nd-preimage resistance and may divide the definition of a collision resistant hash function into two pieces one of which is the weakly collision resistant hash function (WCFHF) that has only 2nd-preimage resistance and the other one is the strongly collision resistant hash function (SCFHF) that is defined only to have collision resistance, [4].

\noindent \textbf{Corollary 2.1.5} A strongly collision free hash function is also a weakly collision free hash function.

Even collision resistance implies 2nd-preimage resistance as above corollary states, we do not have an implication that collision resistance implies preimage resistance.  The following example illustrates this.

\noindent \textbf{Example 2.1.6} Let f be a CRHF, and define a new function h as
\begin{displaymath}
h(x) = \left\{ \begin{array}{ll}
x, & \textrm{$|x|=mn$, $m$ is a nonnegative integer.}\\
f(x), & \textrm{else.}\\
\end{array} \right.
\end{displaymath}

For the above example h is just the identity map for $|x|=mn$, so has no collisions for this case.  For $|x| \neq mn$ case, h is equal to $f(x)$, so finding a collision for h is as hard as finding a collision for f(x).  Therefore, h has collision resistance property.  Now take any $x\ni |x| = mn$, then preimage of $x$ is trivially $x$ itself.  Hence h has not the property of preimage resistance.

Even though collision resistance does not imply preimage resistance, with a weak assumption on the relative cardinalities of the domain and range of the collision resistant hash function, one can prove that collision resistance does now imply preimage resistance.  [4] proves this with the following theorem.

\noindent \textbf{Theorem 2.1.7} Suppose $h:X \rightarrow Z$ is a hash function where $|X|$ and $|Z|$ are finite and
$|X| \geq 2|Z|$.  Suppose $A$ is an inversion algorithm for $h$.  Then there exists a propbabilistic Las Vegas algorithm which finds a collision for $h$ with probability at least $1/2$.

Even though the above theorem is for domains of finite size, the argument is also valid for the infinite domain $\{0,1\}^*$.

For keyed hash functions, the most important group is the message authentication codes (MACs) which are used for message authentication.  The formal definition of a MAC is as follows, [3].

\noindent \textbf{Definition 2.1.8} A \textbf{message authentication code} algorithm is a family of functions $h_k:\{0,1\}^*\rightarrow\{0,1\}^n$ parameterized by a secret key $k\in\{0,1\}^\kappa$, with the following properties:

1. For any $k$ and $x$, $h_k(x)$ is easy to compute.

2. Given zero or more $(x_i,h_k(x_i))$ pairs it is hard to compute any $(x, h_k(x))$ pair for $x\neq x_i$.  This property is called as computation resistance.

\noindent \textbf{Corollary 2.1.9} A MAC algorithm is preimage resistant, 2nd-preimage resistant, and collision resistant for people not knowing the key.

\noindent \textbf{Corollary 2.1.10} Computation resistance implies key non-recovery.
\subsection{Basic Security Considerations}
The reason that one needs preimage, 2nd-preimage, or collision resistance for a hash function is related to ensure the security of the application processed.  For instance suppose that we have a data integrity and message authentication
scheme as described in the intoruction part of this paper.  Now suppose that there is an adversary $A$ on the communication
link of $S$ and $R$.  If $A$ computes a signature on a message digest $z$ and finds an $x \ni z = h(x)$, then
$(x, y=E_K(z=h(x)))$ becomes a valid forgery.  Thus a secure hash function should be preimage resistant.  Now suppose that $S$ sends $(x, y=E_K(h(x)))$ to $R$.  During this transmission if $A$ finds an $x' \ni h(x') = h(x)$, then
$(x', y=E_K(h(x)))$ is a valid forgery.  Hence, a good hash function should be 2nd-preimage resistant.  Finally, suppose that $A$ finds two distinct $x, x'\ni h(x) = h(x')$ and persuades $S$ to sign $h(x)$, then $(x', y=E_K(h(x)))$ is again a valid forgery.  Therefore, a secure hash function should also be collision resistant.

For a MAC, a forgery means to break the computation resistance property in some way.  By Corollary 2.1.10, this can be done by recovering the secret key.  However, key recovery is not a necessary condition for a MAC forgery though it is a sufficient condition.  Sometimes, such a forgery can be done by using only zero or more $(x, h_k(x))$ pairs open to adversary $A$.  Some sources classify attacks on a MAC algorithm with respect to the $A$'s ability to control known
$(x, h_k(x))$ pairs, as [3] does as follows.

1. \textbf{Known-Text Attack}: One or more $(x_i, h_k(x_i))$ pairs are already available to $A$.

2. \textbf{Chosen-Text Attack}: One or more $(x_i, h_k(x_i))$ pairs are available to $A$, where $x_i$'s are chosen by $A$ independently.

3. \textbf{Adaptive Chosen-Text Attack}: One or more $(x_i, h_k(x_i))$ are pairs available to $A$, where $x_i$'s can be chosen by $A$ successively, based on the results of prior queries.

When an $A$ forges a hash function, there are two possibilites related to his control on the fake message he constructed.  If he has partial or full control over the fake message, he is said to make a selective forgery.  On the other hand, if he is only able to contruct a fake message but has no control on the fake message, he is said to make an existential forgery.
\newpage
\section{Block-Cipher Based Hash Functions}
As we stated in Section 2, we can divide hash functions as block-cipher and non-block-cipher based hash functions.  In this section, we will explore the block-cipher based hash functions, by first introducing what a block cipher is.  [5] defines block ciphers mathematically as follows.
\subsection{Block-Ciphers}
\noindent \textbf{Definition 3.1.1} Let $\kappa$ and $l$ be two strictly positive integers.  A \textbf{finite pseudorandom permutation (PRP)} or a \textbf{block-cipher}, with key length $\kappa$ and block length $l$ is a function $F:\{0,1\}^{\kappa}\times\{0,1\}^l\rightarrow\{0,1\}^l$ where $F(K,.)$ is a permutation for each $K\in\{0,1\}^{\kappa}$.

The following definiton implicitly gives the notion of the security of a block cipher.

\noindent \textbf{Definition 3.1.2} Let $D$ be a PRP adversary and let $F$ be a PRP with block length $l$ and key length
$\kappa$.  Define the \textbf{advantage} of $D$ as follows:
\begin{displaymath}
Adv_D^{PRP}(F) = Pr[K \leftarrow \{ 0,1 \}^{\kappa}:D^{F_K(.)}=1]-Pr[\pi \leftarrow Perm_l:D^{\pi(.)}=1],
\end{displaymath}
where $Perm_l$ is the set of all permutations from ${\{ 0,1 \}^l}$ to ${\{ 0,1 \}^l}$.

The meaning of the above definition is the following.  $D$ makes some queries for some amount of time and outputs a $'1'$ bit indicating that he believes he is given a PRP.  The two probabilities are according to whether the adversary is given a PRP or a random permutation.

Now, how secure a block cipher is defined as follows.

\noindent \textbf{Definition 3.1.3} Let $\kappa$ and $l$ be two strictly positive integers.  Let $F$ be a block cipher.  We say adversary $D$ can $t,q,\epsilon-$ \textbf{distinguish} $F$ \textbf{from a random permutation} if $D$ runs at most $t$ steps, makes at most $q$ queries and $Adv_D^{PRP}(F)\ge \epsilon$.

\noindent \textbf{Example 3.1.4} Here is an example of an insecure block cipher.  Let $\kappa=l=64$ and $X$ be a PRP defined as $X_K(M) = M \bigoplus K$.  Now suppose that we are able to make two queries.  So, we first query as $X_K(0^{64}) = 0^{64} \bigoplus K = K$, hence we got the key if we are given the PRP.  Now we make the query $X_K(1^{64})$.  If we have $1^{64}\bigoplus K$, we output a $'1'$ bit that shows we believe that we were given the PRP.  For a random permutation to have the property $\pi(1^{64})=\pi(0^{64})\bigoplus1$ has the probability $1/{(2^{64}-1)}$.  Hence we compute
$Adv_D^{PRP}(X) = 1-1/({2^{64}-1)}$.
Hence, we conclude that a secure block cipher should have small $\epsilon$ even we have big $q$ and $t$ in the sense of above definitions.
\subsection{Types of Block-Cipher Based Hash Functions} In this section we will examine some examples of block-cipher based hash algorithms.  In general hash algorithms have an iterative nature.  Therefore, it is useful to make the following definition ([3]), to classify hash functions based on the amount of their block-cipher operations.

\noindent \textbf{Definition 3.2.1} Let $h$ be an iterated hash function constructed from a block cipher $E$ which performs $s$ block encryptions to process each message block.  Then the \textbf{rate} of $h$ is $1/s$.

Now we will give some examples of block-cipher based hash functions and indicate to which attacks they are open, though we will later in this paper examine the well-known attack methods.

\noindent \textbf{1. Rabin's Hash}

\noindent This hash scheme is due to [6]. First the message is divided into $t$ blocks whose lengths are equal to the block length of the PRP $E$.  Then the following, rate-$1$, algorithm gives us a hash value.
\begin{displaymath}
H_0 = IV \textrm{(Initializing Value), }H_i = E(M_i, H_{i-1})\textrm{, }H(M)=H_t\textrm{, where $i\in \{1,...,t\}$}.
\end{displaymath}
The above hash method is insecure and open to birthday attack for small sized hash values as [7] shows.  It is also open to meet-in-the-middle attack, [8].

\noindent \textbf{2. Combined Plaintext-Ciphertext Chaining Hash}

\noindent This rate-$1$ hash scheme was offered by [9], which uses one common secret key for privacy and authentication.  The algorithm is as follows.
\begin{displaymath}
M=M_t...M_1, M_{t+1}=IV, H_i = E(K, M_i\bigoplus M_{i-1}\bigoplus H_{i-1}),
\end{displaymath}
\begin{displaymath}
H(M)=H_{t+1}\textrm{, where $i\in \{1,...,t\}$}.
\end{displaymath}
This algorithm is open to birthday attack, [8] .

\noindent \textbf{3. Key Chaining Hash}

\noindent This rate-1 hash function was constructed by [10] and [11] to strengthen Rabin's hash with the following algorithm.
\begin{displaymath}
H_0 = IV, H_i = E(M_i\bigoplus H_{i-1}, H_{i-1}), H(M)=H_t\textrm{, where $i\in \{1,...,t\}$}.
\end{displaymath}
This algorithm is open to meet-in-the-middle attack.  Even though there have been made many improvements on this scheme, as [12] and [13], it has still some weaknesses.

\noindent \textbf{4. Matyas-Meyer-Oseas Hash and Davies-Meyer Hash}

\noindent These hash algorithms successfully resist to meet-in-the-middle attack because of the one-wayness of the underlying PRP.  However, they have still some weaknesses related to the key used, [14], [15].  The algorithms\footnote{$g$ is a function from $\{0,1\}^l $ to $\{0,1\}^\kappa $, if $\kappa = l$, then $g$ can be chosen as the identity map.} are as follows for Matyas-Meyer-Oseas and Davies-Meyer respectively, [3], [8].
\begin{displaymath}
H_0 = IV, H_i = E_{g(H_{i-1})}(M_i)\bigoplus M_i, H(M)=H_t\textrm{, where $i\in \{1,...,t\}$}.
\end{displaymath}
\begin{displaymath}
H_0 = IV, H_i = E_{x_i}(H_{i-1})\bigoplus H_{i-1}, H(M)=H_t\textrm{, where $i\in \{1,...,t\}$}.
\end{displaymath}
\noindent \textbf{5. Miyaguchi-Preneel Hash}

\noindent This rate-1 hash was proposed by [16].  [17] showed that this algorithm is open to differential cryptanalysis.  The algorithm is defined as follows, [3].

\begin{displaymath}
H_0 = IV, H_i = E_{g(H_{i-1})}(M_i)\bigoplus H_{i-1}\bigoplus M_i, H(M)=H_t\textrm{, where $i\in \{1,...,t\}$}.
\end{displaymath}

We noted that the above hash schemes have all hash values of length equal to the block length of the PRP.  However there are also hash schemes which have hash values of length twice of the block length of the PRP.  The motivation for these hash functions are to thwart birthday attacks by expanding the sample space.  Now we will give some examples of these types of block-cipher based hash functions.

\noindent \textbf{6. Yi-Lam Hash}

\noindent This hash algorithm was proposed by [18].  For this function, we have $\kappa = 2l$ and $|h(M)| = 2l$.  We have the following algorithm for this hash function.

\begin{displaymath}
H_0 = IV_1, G_0 = IV_2,
\end{displaymath}
\begin{displaymath}
K_i = H_{i-1}||G_{i-1},
\end{displaymath}
\begin{displaymath}
H_i = E_{K_i}(M_i)\bigoplus M_{i}, G_i = (E_{K_i}(M_i)\bigoplus G_{i-1})[+]H_{i-1},
\end{displaymath}
\begin{displaymath}
H(M)=H_t||G_t \textrm{, where $i\in \{1,...,t\}$} \textrm{ and } [+] \textrm{ is summation modulo } 2^l.
\end{displaymath}

Despite the fact that this hash scheme was conjectured to be secure against all attacks faster than brute force, it was later proved to be insecure [19], [20].

\noindent \textbf{7. MDC2 and MDC4}

\noindent MDC2 and MDC4 are two modification dedection codes proposed by IBM, and have rates 1/2 and 1/4 respectively, with the algorithms defined in [3].

Now we will examine the most widely used block-cipher based MAC algorithm, CBC MAC.
\subsection{CBC MAC }

\noindent CBC MAC is the mostly used block-cipher based MAC algorithm today.  [21], [22], and [23] give the algorithm of CBC MAC as follows.
\begin{displaymath}
H_0 = IV, H_i = E(K, M_i\bigoplus H_{i-1}), H(M)=H_t, \textrm{where $i\in\{1,...,t\}$}.
\end{displaymath}

[24] proved that CBC MAC is secure for fixed length messages, say for messages of length $ml$ for some $m$.  However, default CBC MAC algorithm is not secure for arbitrary length messages.  Let's give some examples from [3] and [5] to make this point clear.

\noindent \textbf{Example 3.3.1} Let $M=M_1...M_t$ be a message and suppose $(M_1, MAC_1)$ is known.  Now consider the MAC of $MAC_1$ which is $E_k(MAC_1) = E_k(E_k(M_1))$.  Note that this MAC is also the MAC of $M_1||0^l$.  Hence
$(MAC_1, M_1||0^l)$ is a valid existential forgery.

\noindent \textbf{Example 3.3.2} Let $M=M_1...M_t$ be a message and suppose $(M_1, MAC_1)$ and $(M_2, MAC_2)$ are known.  Now consider the MAC of $M_1||N$ which is $E_k(MAC_1\bigoplus N)$, where $N$ is an arbitrarily chosen message block.  Note that this MAC is also the MAC of $M_2||MAC_1\bigoplus N\bigoplus MAC_2$, this again yields to a valid forgery.

\noindent \textbf{Example 3.3.3} Let $M=M_1...M_t$ be a message and suppose $(M_1, MAC_1)$ is known.  Then the MAC of $M_1||(M_1\bigoplus MAC_1)$ is immediately known.

In order to deal with the above deficiency of the CBC MAC, other variants of this algorithm were developed and proved to be secure.  Now, we will give these in the following part of this section, [5].

\noindent \textbf{1. EMAC}

\noindent Suppose that we have a CBC MAC $h^{CBC}(K_1, M)$.  Now we define the EMAC as
$h^{EMAC}(K_1,K_2, M)=E(K_2, h^{CBC}(K_1, M))$ with an additional key $K_2\ne K_1$, [25].  Since the underlying PRP of the MAC hash only takes inputs of block length $l$, the domain of EMAC is $(\{0,1\}^l)^+$.  [26] has showed that the probability of breaking EMAC, $Pr(forge)\le {2\sigma^2}/2^l$, where $\sigma$ is the total number of blocks of messages whose MACs are known.  Clearly, the number of encryptions is ${|M|/l}+1$.

Even though we have expanded the domain of securely hashed messages from $\{0,1\}^{ml}$ to $(\{0,1\}^l)^+$, we are still not able to MAC arbitrary length messages.  The first variant of CBC MAC dealing with this problem is the EMAC$^*$, which is defined as follows.

\noindent \textbf{2. EMAC$^*$}

\noindent Suppose that we have an EMAC, $h^{EMAC}$.  Now we define EMAC$^*$ as
$h^{EMAC^*}(K_1,K_2, M)=h^{EMAC}(K_1,K_2, M||10^{l-1-|M|modl})$.  The disadvantage of this scheme is that it creates an unnecessary block when $|M|=tl$ for some integer t.  The number of encryptions is $\lceil{{(|M|+1)}/l}\rceil+1$.

\noindent \textbf{3. ECBC}

\noindent To deal with the extra padding problem of EMAC$^*$, a new algorithm was described, called ECBC, which does padding only when necessary.  The algorithm of ECBC, $h^{ECBC}$, is defined as follows.
\begin{displaymath}
\textrm{if } M\in (\{0,1\}^l)^+ \textrm{ then return } h^{EMAC}(K_1,K_2, M)
\end{displaymath}
\begin{displaymath}
\textrm{else return } h^{EMAC}(K_1,K_3, M||10^{l-1-|M|modl}).
\end{displaymath}

In the above algorithm , $K_3$ is a key distinct from $K_2$.  Otherwise, there are trivial collisions.  For example, lets take $h^{ECBC}(K_1,K_2,K_3, M)$ and $h^{ECBC}(K_1,K_2,K_3, M||10^{l-1-|M|modl)}$.  They have the same hash value for any $M\notin(\{0,1\}^l)^+$.  The number of encryptions in ECBC is $\lceil{|M|/l}\rceil+1$.  Theorem 3.3.5 proves that ECBC is secure, [27], but first consider the following definiton.

\noindent \textbf{Definition 3.3.4} Let $l,m,m'\ge{1}$, then the \textbf{collision probability} of the CBC MAC, $V_l(m,m')$ is defined as follows.
\begin{displaymath}
V_l(m,m') = \max_{M\in \{0,1\}^{lm},\textrm{ } M'\in\{0,1\}^{lm'},\textrm{ } M\ne M'}
{\{Pr[\pi\leftarrow Perm_l: h^{CBC}(\pi, M)=h^{CBC}(\pi, M')]\}}.
\end{displaymath}
\noindent \textbf{Theorem 3.3.5} Fix $l\ge 1$ and let $N=2^l$.  Let $D$ be an adversary which asks at most $q$ queries each of which is at most $ml$-bits.  Assume $m\le N/4$.  Then
\begin{displaymath}
Pr[\pi_1, \pi_2, \pi_3\leftarrow Perm_l: D^{h^{ECBC}(K_1,K_2,K_3, .)}=1]-Pr[R\leftarrow Rand(\{0,1\}^*, l):D^{R(.)}=1]
\end{displaymath}
\begin{displaymath}
\le \frac{q^2}{2}V_l(m,m)+\frac{q^2}{2N}\le \frac{(2m^2+1)q^2}{N}.
\end{displaymath}
\noindent \textbf{4. FCBC}

\noindent In order to decrease the number of the encryptions in ECBC PRP from $\lceil{|M|/l}\rceil+1$ to $\lceil{|M|/l}\rceil$, another algorithm was constructed, called FCBC, defined as follows.
\begin{displaymath}
\textrm{if } M\in (\{0,1\}^l)^+ \textrm{ then  } K\leftarrow K_2, \textrm{ and } P\leftarrow M
\end{displaymath}
\begin{displaymath}
\textrm{else } K\leftarrow K_3, \textrm{ and } P\leftarrow M||10^{l-1-|M|modl}
\end{displaymath}
\begin{displaymath}
\textrm{ Let } P = P_1...P_m, \textrm{ where } |P_1|=|P_2|=...=|P_m|=l
\end{displaymath}
\begin{displaymath}
C_0=0^n
\end{displaymath}
\begin{displaymath}
\textrm {for } i\leftarrow 1 \textrm{ to }m-1 \textrm{ do}
\end{displaymath}
\begin{displaymath}
C_i\leftarrow E_{K_1}(P_i\bigoplus C_{i-1})
\end{displaymath}
\begin{displaymath}
\textrm{return } E_{K}(P_m\bigoplus C_{m-1})
\end{displaymath}

Theorem 3.3.6, [27], proves the security of the above algorithm.

\noindent \textbf{Theorem 3.3.6} Fix $l\ge 1$ and let $N=2^l$.  Let $D$ be an adversary which asks at most $q$ queries each of which is at most $ml$-bits.  Assume $m\le N/4$.  Then
\begin{displaymath}
Pr[\pi_1, \pi_2, \pi_3\leftarrow Perm_l: D^{h^{FCBC}(K_1,K_2,K_3, .)}=1]-Pr[R\leftarrow Rand(\{0,1\}^*, l):D^{R(.)}=1]
\end{displaymath}
\begin{displaymath}
\le \frac{q^2}{2}V_l(m,m)+\frac{q^2}{2N}\le \frac{(2m^2+1)q^2}{N}.
\end{displaymath}

\noindent \textbf{5. XCBC}

\noindent A latest version of the CBC MAC variants is the XCBC algorithm, which gets rid of using multiple keys as the secret key of the underlying PRP, and hence uses the same key for all PRP operations.  This algorithm is defined as follows.
\begin{displaymath}
\textrm{if } M\in (\{0,1\}^l)^+ \textrm{ then  } K\leftarrow K_2, \textrm{ and } P\leftarrow M
\end{displaymath}
\begin{displaymath}
\textrm{else } K\leftarrow K_3, \textrm{ and } P\leftarrow M||10^{l-1-|M|modl}
\end{displaymath}
\begin{displaymath}
\textrm{ Let } P = P_1...P_m, \textrm{ where } |P_1|=|P_2|=...=|P_m|=l
\end{displaymath}
\begin{displaymath}
C_0=0^n
\end{displaymath}
\begin{displaymath}
\textrm {for } i\leftarrow 1 \textrm{ to }m-1 \textrm{ do}
\end{displaymath}
\begin{displaymath}
C_i\leftarrow E_{K_1}(P_i\bigoplus C_{i-1})
\end{displaymath}
\begin{displaymath}
\textrm{return } E_{K_1}(P_m\bigoplus C_{m-1}\bigoplus K)
\end{displaymath}

Theorem 3.3.7, [27], proves the security of the above algorithm.

\noindent \textbf{Theorem 3.3.7} Fix $l\ge 1$ and let $N=2^l$.  Let $D$ be an adversary which asks at most $q$ queries each of which is at most $ml$-bits.  Assume $m\le N/4$.  Then \newpage
\begin{displaymath}
|Pr[\pi_1\leftarrow Perm_l; K_2, K_3 \leftarrow \{0,1\}^l: D^{h^{FCBC}(\pi_1,K_2,K_3, .)}=1]
\end{displaymath}
\begin{displaymath}
-Pr[R\leftarrow Rand(\{0,1\}^*, l):D^{R(.)}=1]
\end{displaymath}
\begin{displaymath}
\le \frac{q^2}{2}V_l(m,m)+\frac{(2m^2+1)q^2}{N}\le \frac{(4m^2+1)q^2}{N}.
\end{displaymath}
\section{Non-Block-Cipher Based Hash Functions}
In this section, we give a brief review of non-block-cipher based hash functions.  The general property of these functions is that they are preimage resistant and do not use a PRP as a primitive element.  Moreover, they may have mathematically complex structures.
\subsection{Some Special Examples}
Now, we will give information on some special examples of non-block-cipher based hash functions.

\noindent \textbf{1. Hash Functions Using Matrices}

\noindent [28] (Random Matrix Hashing Algorithm) and [29] propose hash functions based on matrix algebra.  In [28], to get the hash value,  a secret key is used which is in the form of a $t\times l$ matrix.  In [29], the hash value $h(M)$ of a message $M$ is computed as $h(M)=M^TRM$, where $R$ is a randomly chosen $t\times t$ matrix.  [29] has been shown to have some weaknesses.

\noindent \textbf{2. Hash Functions Using Number Theory}

\noindent These functions are based on the difficulty of solving problems in number theory, and most of them are constructed with a modular arithmetic operation.  Here are some examples.

\noindent \textbf{2.1  RSA-Cipher Block Chaining}

\noindent The algorithm of this RSA based CBC function is defined as follows, [8].

\begin{displaymath}
H_0 = IV, H_i = (H_{i-1}\bigoplus M_i)^e \textrm{ mod } N, H(M)=H_t\textrm{, where $i\in \{1,...,t\}$}.
\end{displaymath}

The length of $N$ defines the trade off between the security and speed of this algorithm.  In order to acquire a higher speed, this algorithm was modified by replacing $e$ with 2, or applying this squared modulus only on either $H_{i-1}$ or $M_i$, [12], [30].  Similar constructions using squaring can be found in [8], [31], [32], [33], [34].

\noindent \textbf{2.2  Chaum-van Heijst-Pfitzmann Hash Function}

\noindent This hash function is based on a discrete logarithm problem, which is defined as follows, [35].  $p$ and $q=(p-1)/2$ are two big primes and $\alpha$ and $\beta$ are two primitives of $Z_p$.  It is difficult to know the value of $log_{\alpha}\beta$ and one defines the Chaum-van Heijst-Pfitzmann Hash Function $h:\{0,...,q-1\}^2\rightarrow Z_p-\{0\}$ as $h(m_1, m_2) = \alpha^{m_1}\beta^{m_2}\textrm{ mod }p$.  The contrapositive of the following theorem proves the security of the above function.

\noindent \textbf{Theorem 4.1.1} Given one collision for the Chaum-van Heijst-Pfitzmann Hash Function $h$, it is easy to compute $log_{\alpha}\beta$.

\noindent \textbf{3. Other Special Hash Functions}

\noindent There are other hash schemes based on different systems.  For example, there are hash functions based on claw-free permutations ([36]), knapsack problem ([8]), and cellular automata ([37]).
\subsection{MD4 Family}
MD4 family functions were already overviewed well in the appendix part of the survey [1].  Alas, we do not give the details of them here again.  Some of these functions are named MD2, MD4, MD5, HAVAL, Snefru-8, RIPEMD 128, SHA-1, RIPEMD 160, MAA, DSA, BCA, FFT-Hash I, FFT-Hash II, N-Hash.  The general property of these functions are they are fast on 32-bit machines and most of them are considered to be secure.  SHA-1 is one of the most used standards today, [41].    
\section {Constructing Hash Functions}
\noindent It has been shown that we can construct a SCFHF with an
infinite domain from a SCFHF with a finite domain.  The following
two theorems summarize this, with details in [4].

\noindent \textbf{Theorem 5.1} Suppose $h:(Z_2)^m\rightarrow (Z_2)^t$ is a
SCFHF, where $m\ge t+2$.  Then one can construct a function
\begin{displaymath}
h^*:\bigcup_{i=m}^\infty (Z_2)^i\rightarrow (Z_2)^t, \textrm{which
is a SCFHF}.
\end{displaymath}
\noindent \textbf{Theorem 5.2} Suppose $h:(Z_2)^{t+1}\rightarrow (Z_2)^t$ is a
SCFHF.  Then one can construct a function
\begin{displaymath}
h^*:\bigcup_{i=t+1}^\infty (Z_2)^i\rightarrow (Z_2)^t, \textrm{which
is a SCFHF}.
\end{displaymath}

\section {Attack Methods}
\noindent Throughout the paper, we gave some notions of the security of hash functions in many parts.  Therefore, this section is indeed a complementary part to all of those information.  In this section we will talk about the possible methods of attacks on hash functions.

\noindent In  general a weak hash function means that its algorithm is open to a kind of attack, possibly to many.  For instance, the simplest possible attack is, for a message $M$, to find a correct hash value $MD = h(M)$ with a random guess, which has the probability $1/2^l$. Here are the other possible attacks on hash functions.

\noindent \textbf{1. Birthday Attack}

\noindent This attack is based on the famous Birthday Paradox\footnote{This is indeed a mathematical fact, not a paradox.}, which is defined as follows.  Let $f:\{p_1,p_2,...\}\rightarrow\{d_1, ...,d_{365}\}$ be a function from the set of all people ($p_i'$s) to the days of the year ($d_j'$s).  $f$ does simply give the birthday of the person it was applied.  Then finding two people, $p_i$ and $p_k$ with probability $P$ such that $f(p_i)=f(p_k)$, requires almost $\sqrt{nln(1/1-P)}$ applications of this function.  The same argument can be applied to any hash function to find collisions.  However, if one chooses a set of hash values with sufficiently big length, this attack does not make any sense from the point of computational feasibility.

\noindent \textbf{2. Key Search and Pseudo Attacks}

\noindent Key search attack is used for the keyed hash functions to make a key recovery, hence to break the hash algorithm.  This attack is defined as follows.  Suppose there is one or more $(M_i, h(K,M_i))$ pairs known to $A$.  Then $A$ selects arbitrary elements, $K_i'$s from the key space $\{0,1\}^\kappa$ and tests whether $h(K_i,M_i) = h(K,M_i)$ or not.  Finally, $A$ tries to determine a suitable secret key.  [38] makes a good analysis of the limits of this kind of key search technique.  In the pseudo attack, again a key is tried to be determined, but this time it is sufficient for the key to work only with the known $(M_i, h(K,M_i))$ pairs.  This may cause to some weaknesses.  These techniques are again subject to computational feasibility.

\noindent \textbf{3. Meet-in-the-middle Attack}

\noindent Meet-in-the-middle attack is originated from the birthday attack.  This attack is used in the iterated hash schemes to break preimage or 2nd-preimage resistance.  The attack is defined as follows, [8].  First a selected message is divided into two parts.  One starts from the initial value and goes forward, at the same time he starts from the hash result and comes backward.  The probability of getting a collision in the intermediate stage is the same probability in the birthday attack.  For details, one can look at [38], [39].

\noindent \textbf{4. Correcting Block Attack}

\noindent This attack is defined as follows.  For a given hash value $MD$, one selects a message $M$ and starts to concatenating a redundancy to this message until the hash value $h(M)$ will be equal to $MD$.

\noindent \textbf{5. Differential Cryptanalysis}

\noindent This attack is due to [40], which is done by examining the correlation between the inputs and outputs of a hash function.

\noindent \textbf{6. Fixed Point Attack}

\noindent This attack is again for the iterated hash schemes, which aims to find a fixed point in an intermediate stage.  For instance suppose, we have a round function $f$, which satisfies at an intermediate stage $i$ the equation $f(x_i,m_i)=x_i$.  Then we may replace $m_i$ with an other block $m'_i$ and can get the same hash value as a result. 

\section {Conclusion}
We intended to give an overview of almost all types of hash functions constructed during the short history of hash algorithms.  We included mathematical definitions, theorems and facts to make the issue more precise.  We gave information on sufficient security levels for hash functions by introducing possible attack methods on these functions.  Hence, a general reader may now use this technical report as a survey of hash functions.
\section{Notes}
\noindent This paper has not been published yet, the reference to this paper can be currently done as $'$T. Ozsari. A Hash of Hash Functions. Technical Report. Departments of Mathematics and Computer Engineering, Koç university, October 2003.$'$

\newpage

\newpage
\begin{thebibliography}{99}
\bibitem{1} S. Bakhtiari and R. Savafavi-Naini, J. Pieprzyk. Cryptographic Hash Functions: A Survey.
Technical Report 95-09, Department of Computer Science, University of Wollongong, July 1995.
\bibitem{2} Webster's Internet Ready Dictionary, Version 1.0. Exceller Software Corporation, 1996.
\bibitem{3} A. Menezes, P. C. van Oorschot, S. A. Vanstone. \textit{Handbook of Applied Cryptography}. CRC Press, 1996.
\bibitem{4} D. R. Stinson. \textit{Cryptography: Theory and Practice}. CRC Press, 1995.
\bibitem{5} J. R. Black, Jr. \textit{Message Authentication Codes}. PhD Thesis, University of California, Davis, September 2000.
\bibitem{6} M. O. Rabin. Digitalized Signatures. In R. A. Demillo, D. P. Dopkin, A. K. Jones, and R. J. Lipton, editors, \textit{Foundations of Secure Computation}, pages 155-166, New York, 1978. Academic Press.
\bibitem{7} G. Yuval. How to Swindle Rabin. \textit{Cryptologia}, 3:187-189, July, 1979.
\bibitem{8} J. Pieprzyk and B. Sadeghiyan. \textit{Design of Hashing Algorithms}, Springer-Verlag, 1993.
\bibitem{9} C. H. Meyer and S. M. Matyas. \textit{Cryptography: a New Dimension in Data Security}. Wiley \& Sons, 1982.
\bibitem{10} D. W. Davies. Applying the RSA Digital Signature to Electronic Mail, 1983.
\bibitem{11} D. E. Denning. Digital Signatures with RSA and Other Public-Key Cryptosystems. \textit{Communications of the ACM}, 27(4): 388-392, 1984.
\bibitem{12} D. W. Davies. and W. L. Price. Digital Signaure - an update. In \textit{Proceedings of the Seventh International Conference on Computer Communication}, pages 845-849, 1984.
\bibitem{13} J. Seberry and J. Pieprzyk. \textit{Cryptography, An Introduction to Computer Security}. Prentice Hall, 1989.
\bibitem{14} B. Preneel, R. Govaerts, and J. Vandewalle.  Collision resistant hash functions based on block ciphers. submitted to CRYPTO '91.
\bibitem{15} J. J. Quisquater and J. P. Delescaille. How Easy is Collision Search? New results and applications to DES. In \textit{Advances in Cryptology-CRYPTO '89}, volume 435 of \textit{Lecture Notes in Computer Science}, pages 408-413. Springer-Verlag, 1989.
\bibitem{16} S. Miyaguchi, M. Iwata, and K. Ohta. New 128-bit Hash Function. In \textit{Proceedings of 4th International Joint Workshop on Computer and Communications}, pages 279-288, 1989.
\bibitem{17} E. Biham and A. Shamir. Differential Cryptanalysis of FEAL and N-Hash. In
\textit{Abstracts of EUROCRYPT '91}, pages 1-8, 1991.
\bibitem{18} X. Yi and K.-Y. Lam. A new hash function based on block cipher. \textit{ACISP '97}, Second Australasian Conference on Information Security and Privacy, Springer, LNCS 1270.
\bibitem{19} D. Wagner. Cryptanalysis of the Yi-Lam hash. \textit{Lecture Notes in Computer Science}, vol.1976, 2000.
\bibitem{20} K. Martin. Analysis of a Hash Function of Yi and Lam. \textit{Electronic Letters}, vol.34, no.24, 1998, pp.2327-2328.
\bibitem{21} FIPS 113. Computer data authentication. Federal Information Processing Standards Publication 113, U.S. Department of Commerce/National Bureau of Standards, National Technical Information Service, Spring.eld, Virginia, 1994.
\bibitem{22} Iso/Iec 9797-1. Information technology – security techniques – data integrity mechanism using a cryptographic check function employing a block cipher algorithm. International Organization for Standards, Geneva, Switzerland, 1999. Second edition.
\bibitem{23} Financial Institution Message Authentication (Wholesale), ANSI X9.9-1986, American National Standards Institute.
\bibitem{24} Bellare, M., Kilian, J., and Rogaway, P. The security of the cipher block chaining message authentication code. See www.cs.ucdavis.edu/~rogaway. Older version appears in \textit{Advances in Cryptology – CRYPTO '94} (1994), vol. 839 of \textit{Lecture Notes in Computer Science}, Springer-Verlag, pp. 341–358.
\bibitem{25} Berendschot, A., den Boer, B., Boly, J., Bosselaers, A., Brandt, J.,
Chaum, D., Damgard, I., Dichtl, M., Fumy, W., van der Ham, M., Jansen,
C., Landrock, P., Preneel, B., Roelofsen, G., de Rooij, P., and Vandewalle,
J. \textit{Final Report of Race Integrity Primitives}, vol. 1007 of \textit{Lecture Notes in
Computer Science}. Springer-Verlag, 1995.
\bibitem{26} Petrank, E., and Rackoff, C. CBC MAC for real-time data sources, 1998. Available
from http://philby.ucsd.edu/cryptolib.
\bibitem{27} J. Black and P. Rogaway. CBC MACs for arbitrary length messages: The three key constructions. \textit{Advances in Cryptology-Crypto 2000}, \textit{Lecture Notes in Computer Science}, vol. 1880, Springer-Verlag, Mihir Bellare, editor, pp. 197-215, 2000.
\bibitem{28} B. Banieqbal and S. Hilditch. The Random Matrix Hashing Algorithm. Technical Report UMCS-90-9-1, Department of Computer Science, University of Manchester, 1990.
\bibitem{29} S. Harari. Non-Linear, Non-Commutative Functions for Data Integrity. In \textit{Advances in Cryptology, Proceedings of EUROCRYPT '84}, pages 25-32, 1985.
\bibitem{30} M. Girault. Hash-Functions Using Modulo-N Operations. In \textit{Advances in Cryptology-EUROCRYPT '87}, volume 304 of \textit{Lecture Notes in Computer Science}, pages 218-226. Springer Verlag, 1987.
\bibitem{31} R. R. Jueneman, S. M. Matyas, and C. H. Meyer. Message Authentication. \textit{IEEE Communication Magazine}, 23(9):29-40, 1985.
\bibitem{32} R. R. Jueneman. A high Speed Manipulation Detection Code. In \textit{Advances in Cryptology-CRYPTO '86}, volume 263 of \textit{Lecture Notes in Computer Science}, pages 327-347. Springer-Verlag, 1986.
\bibitem{33} R. R. Jueneman. Electronic Document Authentication. \textit{IEEE Network Magazine}, 1(2):17-23, 1987.
\bibitem{34} I. B. Damgard. A Design Principle for Hash Functions. In \textit{Advances in Cryptology-CRYPTO '89}, volume 435 of \textit{Lecture Notes in Computer Science}, pages 416-427. Springer-Verlag, 1989.
\bibitem{35} D. Chaum, E. Van Heijst, and B. Pfitzmann. Cryptographically strong undeniable signatures, unconditionally secure for the signer. \textit{Advances in Cryptology-CRYPTO '91}, volume 576 of \textit{Lecture Notes in Computer Science}, pages 470-484. Springer-Verlag, 1992.
\bibitem{36} I. B. Damgard. Collision Free Hash Functions and Public Key Signature Schemes. In \textit{Advances in Cryptology-CRYPTO '87}, volume 304 of \textit{Lecture Notes in Computer Science}, pages 203-216. Springer-Verlag, 1987.
\bibitem{37} J. Daemen, R. Govaerts, and J. Vandewalle. A Framework for the Design of One-Way Hash Functions Including Cryptanalysis of Damgard's One-way Function Based on a Cellular Automaton, 1991. In Abstracts of ASIACRYPT '91.
\bibitem{38} B. Preneel. \textit{Analysis and Design of Cryptographic Hash Functions}. Phd thesis, Katholieke University Leuven, January 1993.
\bibitem{39} K. Nishimura and M. Sibuya. Probability To Meet in the Middle.  \textit{Journal of Cryptology} 2:13-22, 1990.
\bibitem{40} E. Biham and A. Shamir. Differential Cryptanalysis of DES-like Cryptosystems.
\bibitem{41} FIPS 180-1, 'Secure Hash Standard', Federal Information Processing Standards Publication 180-1, U.S. Department of Commerce/N.I.S.T., National Technical Information Service, Springfields, Virginia, April 17 1995 (supersedes FIPS PUB 180).
\textit{Abstracts of EUROCRYPT '90}, pages 1-19, 1991.
\end{thebibliography}
\end{document}